# Dipole glass and mixed ferroglass phase in SrTiO$_3$ ceramics doped with manganese

M. D. Glinchuk, P. I. Bykov

*Institute for Problems of Materials Science, NASc of Ukraine, Krjijanovskogo 3, 03680 Kiev 142, Ukraine*

Incipient ferroelectrics with off-center ions like KTaO$_3$:Li attract much attention of the scientists because of their properties anomalies related to the phase transitions induced by the impurities. Recently new incipient ferroelectrics with off-center ions SrTiO$_3$ with Mn$^{2+}$ substituted for Sr$^{2+}$ was revealed (Tkach et. al PRB 73, 104113 (2006). Experimental investigation of Sr$_{1-x}$Mn$_x$TiO$_3$ (0 ≤ x ≤ 0,15) properties, namely temperature, frequency and external electric field dependence of dielectric permittivity at different concentrations gave evidence about phase transitions induced by the impurities. However their nature was not found out up to now because of the absence of the theoretical analysis of the results.

In this paper we performed such analysis by the theoretical description of the temperature of dielectric permittivity maxima dependence on Mn$^{2+}$ concentration, the change of residual polarization with temperature, frequency and temperature dependence of the permittivity with Arrhenius and Vogel-Fulcher law at smaller and larger concentration respectively. The analytical form of nonlinear on d.c. electric field dependence of the susceptibility was also derived and compared with experimental data. The obtained results had shown that at x < 0,03 and 0,15 ≥ x ≥ 0,03 the considered system Sr$_{1-x}$Mn$_x$TiO$_3$ is in dipole glass state and mixed ferroglass phase respectively.

The directions of further investigations of the material which will allow to obtain additional valuable information on physical nature of the observed phenomena are proposed.

## 1. Introduction

Over last decades there has been essential interest in studies of the properties of incipient ferroelectrics doped with the impurities [1]. The most important phenomenon was shown to be the phase transitions induced by the impurities with electric dipole moment originated from off-central position of the impurity in the lattice e.g. Li$^+$ substituted for K$^+$ in KTaO$_3$. Since pure incipient ferroelectrics are known to be in the paraelectric phase up to zero kelvins, the doped systems can be considered as paraelectrics with electric dipoles. Similarly to paramagnetics with impurity magnetic dipoles (see e.g. [2]) all the properties depend on the dipoles concentration, including the sequence of phase transitions with concentration increase from dipole glass DG (spin glass) to ferro-glass phase FG (ferro-spin glass) and finally to ferroelectric FE (ferromagnetic) phase at high enough concentration [3]. Since FG phase is so called mixed phase with coexistence of polar short- and long-range orders which are the characteristics of DG and FE phase respectively, the FG phase properties contain a lot of anomalies including the relaxor characteristics. The model incipient ferroelectric with off-center impurities KLT — K$_{1-x}$Li$_x$TaO$_3$ (0≤ x≤ 0,1) was investigated in many details [1] and it was shown that critical concentration which separates DG and FG phases is equal 0,022, so that x≤ 0,022 and 0,022< x≤ 0,05 respectively are the regions of DG and FG existence. Up to recently there was no analogous system on the basis of another incipient ferroelectric SrTiO$_3$. Recently it was shown that in Mn-doped KTaO$_3$ [4] and in Ca-doped SrTiO$_3$ [5] dopants substitute for A-ion of perovskite structure at an off-center position and induce polar properties at low temperature region.

In the last two years polar dielectric anomaly, with corresponding hysteretic response in polarisation with high dielectric permittivity was observed in Sr$_{1-x}$Mn$_x$TiO$_3$ system (SMT) [6]. In contrast to SMT Mn doping does not induce any anomaly and just reduces the dielectric permittivity value in SrTi$_{1-x}$Mn$_x$O$_3$ (STM). Recently it was shown by ESR measurements [7] that the reason of this different behaviour is off-center position of Mn$^{2+}$ substituted for Sr$^{2+}$ because of large difference of the ions ionic radii. So, SMT system can be considered as the analogous system to the model disordered system KLT, with the similar properties anomalies. Indeed, the diffuse low temperature peak in dielectric permittivity $\varepsilon'(T)$ shifts to higher temperature with increasing of measurement frequency and the amount of off-center ions. The measurements of the dielectric relaxation in SMT in broad frequency range had shown [6], that Arrhenius law and Vogel-Fulcher law describe the observed data in the sample with lower and higher $Mn^{2+}$ concentrations respectively similarly to what was observed in KLT [8]. The dependence of nonlinear susceptibility on d.c. electric field and temperature was similar to what was observed earlier in $KTaO_3$ doped with another off-center ion Nb [9]. Although these similarities were pretty obvious there are no discussion of them for SMT and especially the physical reasons which lay in their background. The latter seems to be important for the description of the observed in SMT properties anomalies and understanding of phase transformations with $Mn^{2+}$ concentration change. In this paper we performed such consideration using the theoretical approach developed by us earlier for KLT and KTN incipient ferroelectrics with off-center impurities [1,3,8,9]. Note, that because of structural phase transition at $T_0= 105K$ in $SrTiO_3$ its symmetry is tetragonal at $T< T_0$ contrary to $KTaO_3$ which conserves cubic symmetry up to zero kelvins. Keeping in mind that $SrTiO_3$ at $T< 105K$ conserves inversion center, has high polarizability, soft mode, high dielectric permittivity and other properties similar to $KTaO_3$, we neglected the small tetragonal distortions which have to be smeared out essentially in the ceramic samples.

## 2. Residual polarization in DG state

Hysteresis loops $P(E)$ in $Sr_{1-x}Mn_xTiO_3$ ceramic at $x= 0,005$ and $x=0,02$ were observed at low temperatures region $T< 80K$ [10]. The general view of hysteresis loops are similar to what was observed many years ago in classical dipole glass KCl:Li at low temperatures [11]. The observed loops in SMT allowed to extract the temperature dependence of the residual polarization for the considered two concentrations of $Mn^{2+}$. The polarization was larger in the sample with larger concentration and it decreases with the temperature increase. The origin of residual polarization is dipole-dipole interaction, that in pair approximation can be written in the form:

$$V_{dd} = \frac{1}{2} \sum_{i,j,\alpha,\beta} K^{\alpha,\beta}(r_{ij}) d^*_{i\alpha} d^*_{j\beta} \qquad \alpha,\beta = x,y,z \qquad (1)$$

Here $i,j$ numerates the dipoles, $d^* = \gamma d(\varepsilon_0 + 2/3)$ is effective dipole moment, $\gamma$ — Lorentz factor, $\varepsilon_0$ — static dielectric permittivity of incipient ferroelectric (host lattice), $d$ is intrinsic dipole moment, determined by the off-center displacement magnitude. The form of $K^{\alpha,\beta}(r_{i,j})$ takes into account direct ($K(r_{ij}) \sim \pm 1/r_{ij}^3 \varepsilon_0$) and indirect via soft optic mode interactions ($K(r_{ij}) \sim \left(1/r_c^2 r_{ij} \varepsilon_0\right) e^{-r_{ij}/r_c}$, $r_c$ is correlation radius) the latter being especially important for high polarizable systems (the detailed form of $K(r_{ij})$ one can find in [1]). It is obvious from Eq.(1), that the reorientation of every dipole in the pair requires surmounting of the additional potential barrier created by an adjacent dipole. The closer dipoles are to one another, the higher potential barrier and hence the longer the relaxation time of the dipole moment of the pair. In the case when total dipole moment of a pair can be oriented parallel or antiparallel to radius vector $r_{ij}$ the mean frequency $\tilde{\nu}$ of the total moment reorientation is determined by the potential barrier $W = d^{*2}/\varepsilon_0 r^3$ and can be written in the case of weakly polarizable systems as [11]:

$$\tilde{\nu} = \nu e^{-W/kT} \qquad (2)$$

where $\nu$ is the frequency of reorientation of isolated dipoles.

After switching off the external electric field that firstly aligned all the dipoles most of the dipoles depolarise in a time $1/\tilde{\nu}$, the closer the dipoles in the pair the longer the time of depolarisation as follows from Eq.(2). The approximate calculation of residual polarization in the time $t$ with account of the distribution of the distances between dipoles $P_r \sim \int e^{-\tilde{\nu}(W)t} dr$ yields [11]:

$$\frac{P_r}{nd^*} = \frac{2\pi}{3} \frac{nd^{*2}}{\varepsilon_0 kT \ln(\nu t)} \quad (3)$$

The Eq.(3) describes pretty good the observed concentrational, temperature and time dependence of residual polarization in KCl:Li. However for high polarizable systems we are interested in we could use Eq.(3) for qualitive or maybe semiquantitative description of experimental data.

In particular temperature dependence of residual polarization for high polarizable systems where $\varepsilon_0$ contrary to KCl is temperature dependent quantity can be estimated as:

$$P_r(T) \sim \frac{\varepsilon_0^2(T)}{T} \quad (4)$$

Keeping in mind the different form of coefficient $K(r_{ij})$ in dipole-dipole interaction (1) for weakly and highly polarizable systems as it was written at the beginning of this chapter, one can expect another form of barrier, e.g. with substitution of $rr_c^2$ for $r^3$ in $W$. This may lead to another dependence on concentration in expression for $P_r$ supposedly much weaker than $n^2$ and to appearance of multiplier $r_c^2 \sim \varepsilon_0(T)$ in denominator of Eq.(3). The latter will change the temperature dependence of residual polarization:

$$P_r(T) \sim \frac{\varepsilon_0(T)}{T} \quad (5)$$

This dependence is represented by solid lines in Fig.1 along with the squares and circles as the experimental points for residual polarization respectively for the samples with x=0,02 and x=0,005. One can see that $P_r(T)$ in the form of Eq.(5) describes pretty good observed temperature dependence of residual polarization at $T> 20$K. It should be noted, that we took $\varepsilon_0(T)$ dependence for SrTiO$_3$ measured on the same samples [6,10]. We did not try to fit the point at $T= 10$K because there was no value of SrTiO$_3$ dielectric permittivity measured in these works. In any case Eq.(5) fits experimental points better than Eq.(4). The more weaker than the ratio of squared concentrations decrease of $P_r$ value in the sample with x=0,005 comparatively to x=0,02 qualitatively agrees with our expectations for high polarizable systems as it was mentioned above. Therefore the observed slim hysteresis loops and the behaviour of residual polarization speaks in favour of the statement that the SMT samples with concentration of Mn$^{2+}$ ions x≤ 0,02 are in dipole glass state. The investigation of the residual polarization evolution in time at low temperature region will pour additional valuable information on the physical background of residual polarisation.

### 3. Ferro-glass phase in Sr$_{1-x}$Mn$_x$TiO$_3$ with x≥ 0,03

The separation of DG and FG phase one of the most complex problems in any disordered ferroelectrics. In the model doped incipient ferroelectric systems like $K_{1-x}Li_xTaO_3$, $KTa_{1-x}Nb_xO_3$ the main features which could pour light on the problem solution is concentrational dependence of the temperature $T_m$ of dielectric permittivity maximum.

It was shown (see e.g. [9] and ref. therein), that for FG and DG respectively

$$T_m \sim \sqrt{x - x_{cr}} \qquad x > x_{cr} \qquad (6a)$$

$$T_m \sim \sqrt{x} \qquad x < x_{cr} \qquad (6b)$$

In Fig.2 one can see, that for $x_{cr}= 0,03$ Eq.(6a) and (6b) fits pretty good experimental points measured in [6] for $Sr_{1-x}Mn_xTiO_3$ at $x> 0,03$ and $x< 0,03$ respectively. The latter agrees with aboveobtained results that the samples with $x\leq 0,02$ are in DG state.

Another distinguished feature can be obtained from measurements of dynamic dielectric permittivity. Namely the dielectric relaxation was shown to obey Arrhenius or Vogel-Fulcher (V-F) law respectively for DG or FG phase in $K_{1-x}Li_xTaO_3$ (KLT) and $KTa_{1-x}Nb_xO_3$ (KTN) doped incipient ferroelectrics [8]. Relaxation dynamic parameters of SMT compositions obtained in [6] had shown that V-F law described the results of measurements at $x\geq 0,03$ while Arrhenius law fitted the results at $x\leq 0,02$. Therefore all the consideration performed in this and previous chapter lead to the conclusion that $Sr_{1-x}Mn_xTiO_3$ system is in DG or FG phase respectively at $x< 0,03$ or $x\geq 0,03$. It should be noted that in SMT V-F law describes the behaviour even at $x= 0,15$ [6], while in model doped incipient ferroelectrics KTL and KTN normal ferroelectric phase with Arrhenius law appeared at some concentration $x\geq x_0$, $x_0$ value being about $0,05$ for KLT [1]. This can be related to the fact that $MnTiO_3$ has no ferroelectric properties contrary to $LiTaO_3$ and $KNbO_3$ which are known ferroelectric materials [12]. It will be interesting to investigate the dynamic permittivity at $x> 0,15$ to check if V-F law describes the behaviour and to study its parameters dependence on x.

## 4. Nonlinear dielectric permittivity in SMT

Nonlinear dielectric permittivity $\varepsilon_{NL}$ is known to be more sensitive (than linear susceptibility $\varepsilon_L$) to the details of dipoles ordering in the systems. For example the investigation of $\varepsilon_{NL}$ (in dc field) in DG state in the vicinity of freezing temperature $T_g$ is very important to answer the question if DG is truly equilibrium phase with the divergency of $\varepsilon_{NL}$ at $T=T_g$ or whether it is just a metastable state without $\varepsilon_{NL}$ divergency at $T=T_g$, but with long-time (up to infinity) relaxation modes (see [3] and ref. therein). At freezing temperature $T=T_g$ cusp-like anomaly along with conventional maximum at $T=T_m$ appears in $\varepsilon_L$. In ordered ferroelectrics like $BaTiO_3$ maximum of $\varepsilon_L$ corresponds to ferroelectric-paraelectric phase transition and so $T_m= T_c$. The calculations performed in phenomenological Landau theory lead to the divergency of both $\varepsilon_L$ and $\varepsilon_{NL}$ at $T= T_c$. However such mean field approach is not correct for disordered ferroelectrics at least in their DG and FG phases. Because of this the calculations beyond mean field approximations was performed recently [9] to describe $\varepsilon_{NL}$ observed in KTN. The nonlinear effects were known to be large enough in this material because the coefficient of nonlinearity is several times larger in KTN than in KLT. Observed in [13] essential nonlinear dependence of dielectric permittivity on dc external field $E$ allows to suppose that the coefficient of nonlinearity in $Sr_{1-x}Mn_xTiO_3$ is large also. For analysis of experimental data it is convenient to write dielectric permittivity in the form of even powers series [9]:

$$\varepsilon(E,T) = \varepsilon_L(T) - \varepsilon_{NL}^{(1)}(T)E^2 + \varepsilon_{NL}^{(2)}(T)E^4 + ... \qquad (7)$$

Here $\varepsilon_{NL}^{(k)}$ is so-called k-th order nonlinear susceptibility, only even power of E being related to the existence of inversion symmetry in SrTiO$_3$ host lattice.

In Fig.3 we depicted dielectric permittivity calculated in the form of Eq.(7) and shown the experimental data by black points. One can see that Eq.(7) fitted experimental points pretty good. It follows from Fig.3, that at T= 294K dielectric permittivity does not depend on E i.e. $\varepsilon$(294K)= $\varepsilon_L$(294K). The contribution of $\varepsilon_{NL}^{(1)}(T), \varepsilon_{NL}^{(2)}(T)$ increase with temperature decrease, although $\varepsilon_{NL}^{(2)}(T)/\varepsilon_{NL}^{(1)}(T) \approx 10^{-3}$ and so the contribution of $\varepsilon_{NL}^{(2)}(T)$ manifest itself only at large external dc electric field and for smaller fields the strait lines with the slope dependent on temperature can fit experimental points. There is no experimental point at $T=T_g$= 5,8K, so we can not discuss $\varepsilon_{NL}$ behaviour at freezing temperature. The calculations, performed for KTN in [9] had shown that $\varepsilon_{NL}^{(1,2)}(T)$ depends essentially on the distribution function of random electric field parameters, namely on $q = E_0/\sqrt{C}$ ratio, where $E_0$ and C are respectively mean field and half-width of distribution function. Since this ratio defines phase diagram of the disordered systems ($q< 1$, $q\geq 1$, $q>> 1$ correspond to DG, FG and FE phases respectively), let's discuss its possible value in SMT. Keeping in mind, that at x=x$_{cr}$ ratio $q=\sqrt{\pi}$ and $q \sim \sqrt{x}$ at arbitrary x we can estimate q-value at the considered concentration x= 0,05 as $q = \sqrt{5\pi/3} \approx 2,3$. The obtained value confirms the FG phase for SMT at x= 0,05. The investigation of nonlinear permittivity temperature dependence for several concentrations in the range x< 0,03 and x$\geq$ 0,03, including the points $T=T_g$ will be the source of important information about the physical nature of observed phenomena.

Therefore on the basis of available experimental data analysis we can conclude that Sr$_{1-x}$Mn$_x$TiO$_3$ ceramics is in dipole glass state or ferro-glass phase respectively at x< 0,03 or x$\geq$ 0,03.

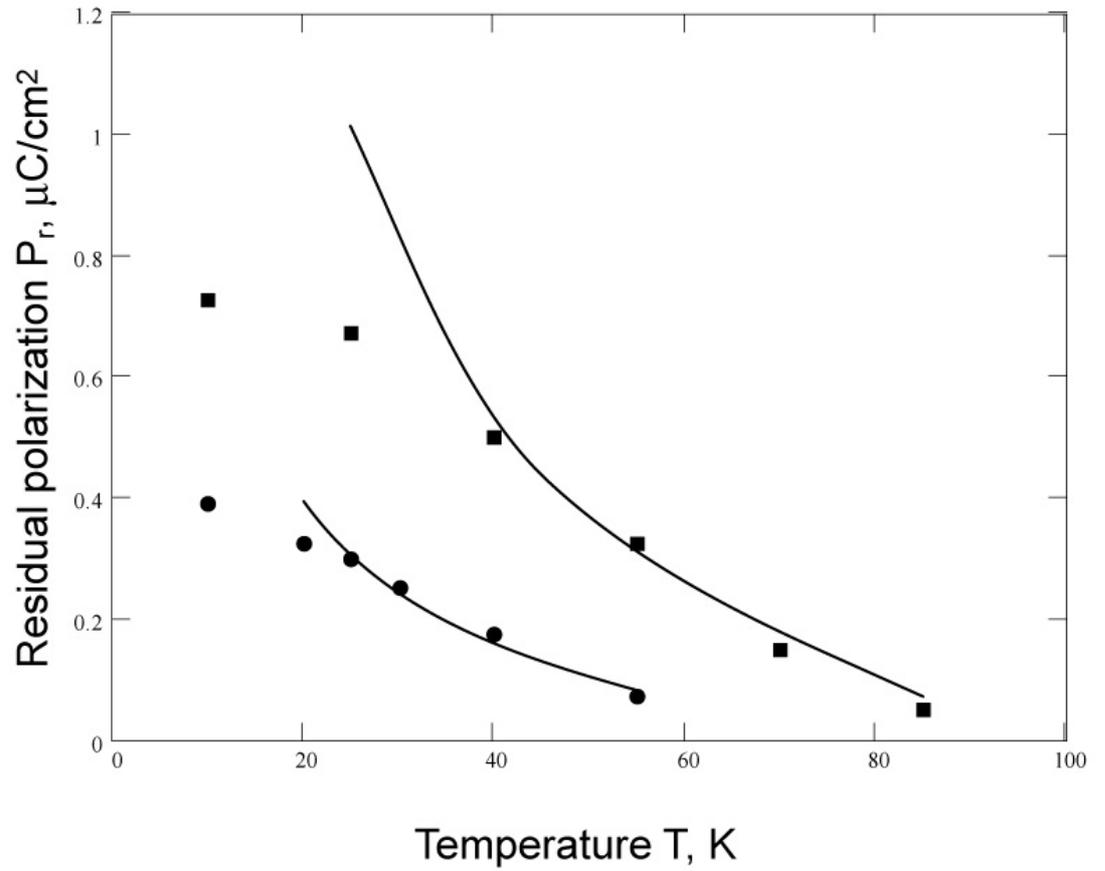

Fig.1 Dependence of residual polarization on temperature for $Sr_{1-x}Mn_xTiO_3$. Experimental data for x= 0,005 (circles) and for x= 0,02 (squares). Solid lines — theory.

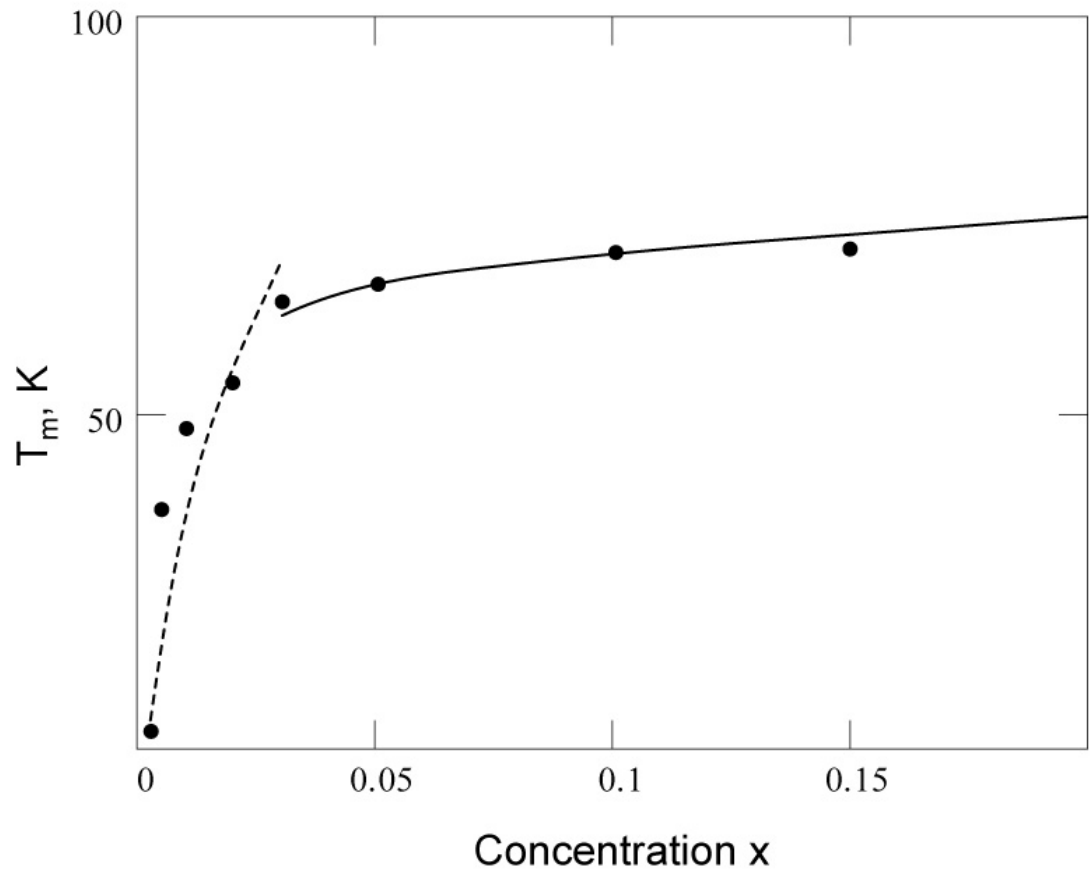

Fig.2 Dependence of temperatures of maximum of dielectric permittivity on concentration of Mn for $Sr_{1-x}Mn_xTiO_3$. Dashed line and solid line are the theory for $x < 0,03$ and $x \geq 0,03$ respectively, black points experimental data.

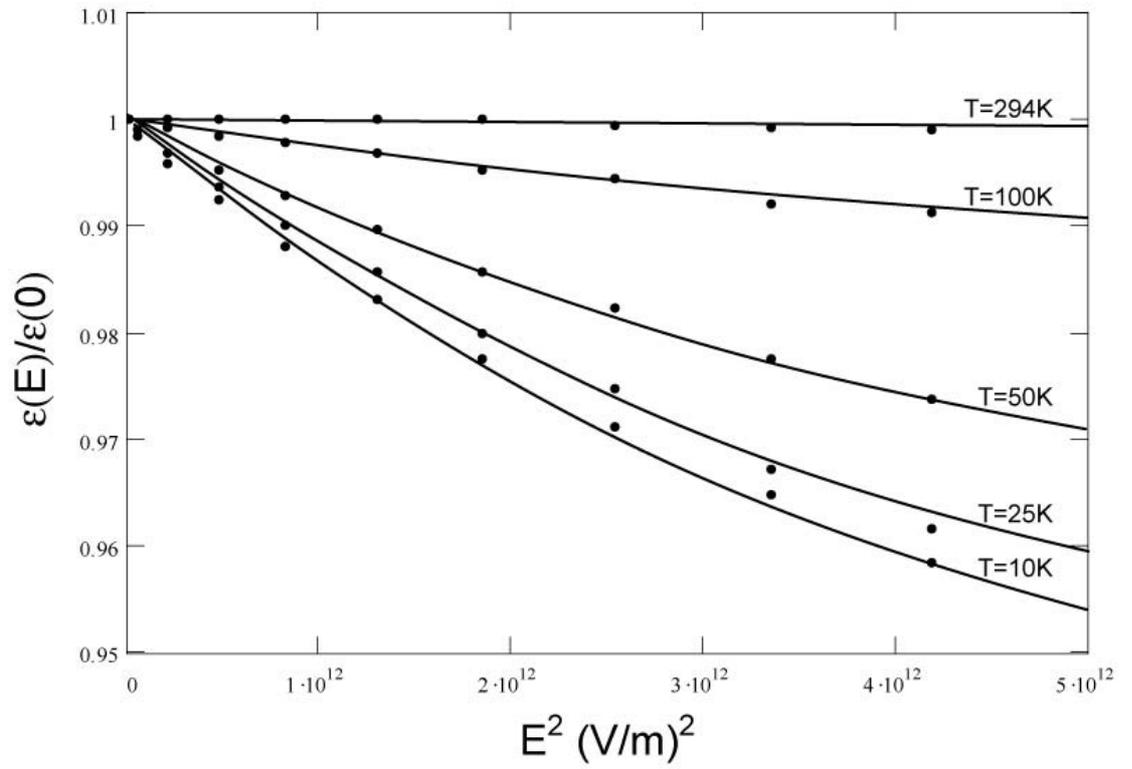

Fig.3 Normalized dielectric susceptibility for $Sr_{1-x}Mn_xTiO_3$, x= 0,05 for different temperatures: points experiment, lines theory (Eq.(7)).